\documentclass[aps,prl,twocolumn,showcaps,superscriptaddress,showpacs,amsmath,amssymb]{revtex4}

\usepackage{graphicx}
\usepackage{dcolumn}
\usepackage{bm}
\usepackage[dvips]{epsfig}
\usepackage{color}
\usepackage{upgreek}
\usepackage{ulem}

\begin{document}

\title{Experimental observation of directional locking and dynamical ordering of colloidal monolayers driven across quasiperiodic substrates }

\author{Thomas Bohlein}
\author{Clemens Bechinger}
\affiliation{2. Physikalisches Institut, Universit\"at Stuttgart, Pfaffenwaldring 57, 70569 Stuttgart, Germany}
\affiliation{Max-Planck-Institut f\"ur Intelligente Systeme, Heisenbergstrasse 3 ,70569 Stuttgart, Germany}


\begin{abstract}
We experimentally investigate the structural behavior of an interacting colloidal monolayer being driven across a decagonal quasiperiodic potential landscape created by an optical interference pattern. When the direction of the driving force is varied, we observe the monolayer to be directionally locked on angles corresponding to the symmetry axes of the underlying potential. At such locking steps we observe a dynamically ordered smectic phase in agreement with recent simulations. We demonstrate, that such dynamical ordering is due to the interaction of particle lanes formed by interstitial and non-interstitial particles.
\end{abstract}


\maketitle


Particles which are driven across periodic substrate potential landscapes show a number of intriguing phenomena. Depending on the direction of the applied driving force $\bf F$,
the orientation of the particle's motion can substantially deviate from $\bf F$ but is locked-in to directions determined by the substrate's symmetry \cite{Korda2002,Roichman2007}.
Examples of such kinetically locked-in states range from atom migration on crystalline surfaces \cite{PL2001}, driven charge density waves \cite{Brown1984} to flux flow in type-II superconductors \cite{Reichhardt1997, Reichhardt1999,Morgan1998}. Also, it has been demonstrated that directional locking can be employed for sorting colloidal particles according to their size,
refractive index or chirality \cite{MacDonald2003,Speer2009}. In contrast to the above examples which have been carried out with diluted systems, only little is known about directional locking in the presence of non-negligible particle interactions. Then, the competition between interparticle forces and those with the substrate leads to complex dynamical ordering phenomena \cite{Reichhartunpublished}. Interestingly, dynamical ordering is not limited to periodic surfaces but is also found on vortex lattices driven across quasiperiodic and disordered pinning sites \cite{Kemmler2006,Silhanek2006,Koshelev1994,Pardo1998}. Recently, directional locking and dynamical ordering was even predicted for interacting colloidal systems on quasiperiodic substrate potentials \cite{Reichhardt2011}. However, both an experimental demonstration and a microscopic understanding of such ordering transitions on quasiperiodic substrates, is still missing.

In this Letter we experimentally demonstrate dynamical ordering of a colloidal monolayer on a quasiperiodic optical interference pattern \cite{Mikhael2008}. When the direction of the driving force is varied with respect to the substrate, we observe directionally locked states with smectic-like order in agreement with recent predictions \cite{Reichhardt2011}. We demonstrate, that this is due to the interaction of particle lanes formed by interstitial and non-interstitial particles. When the angle of {\bf F} deviates from a substrate symmetry direction, the colloidal monolayer partitions into domains which are aligned along different symmetry directions of the substrate.
\begin{figure}[htbp]
  \centering
    \includegraphics[width=0.4\textwidth]{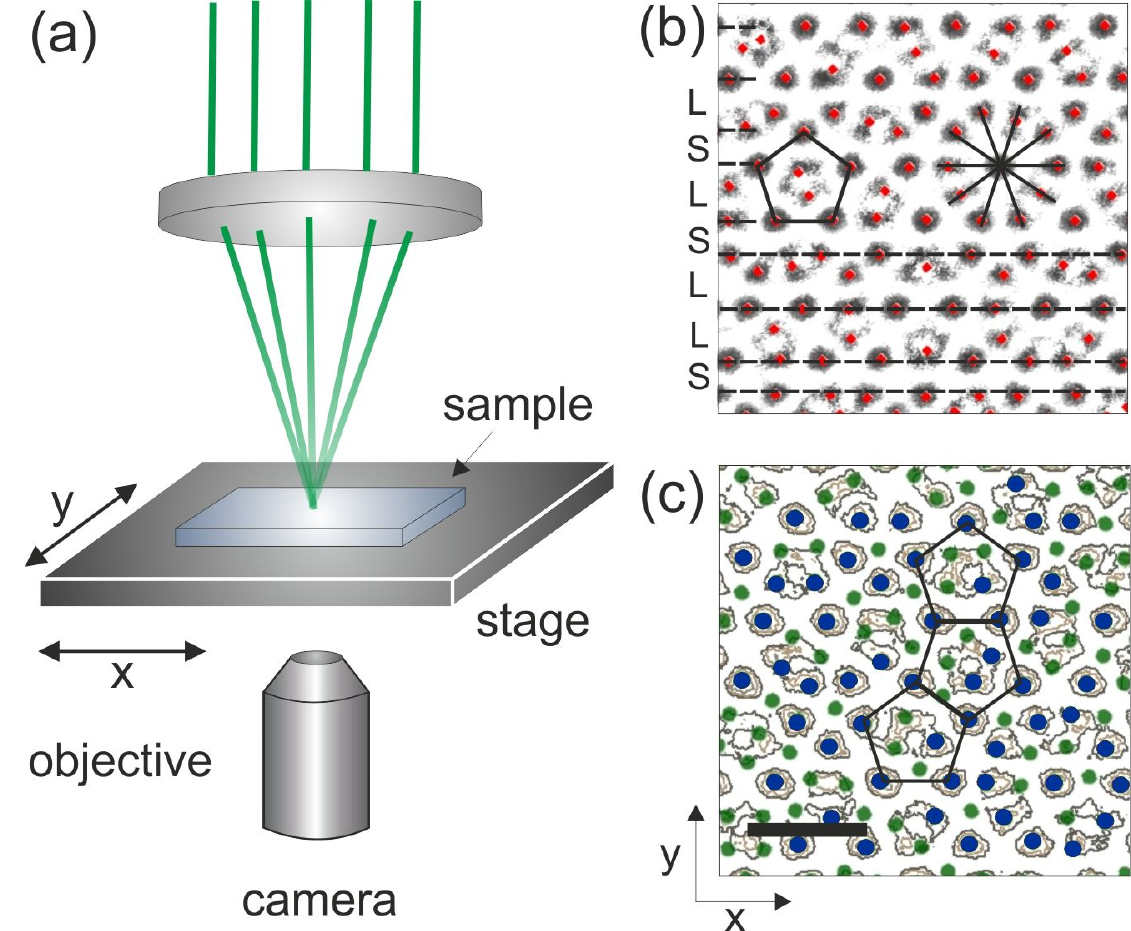}
  \caption{(color online). (a) Five laser beams form a continuous decagonal quasiperiodic potential landscape inside the sample cell which is mounted onto a xy piezo stage for application of lateral viscous forces. (b) Measured intensity distribution with deep potential wells marked in red. The pattern displays decagonal symmetry and the predominating motifs are pentagons and flowers (indicated with solid lines). Deep potential wells are aligned along symmetry directions and form a Fibonacci sequence (dashed lines).
  (c) Particle positions (blue and green dots) superimposed with the quasiperiodic potential for $I = 6.4 \mu W/\mu m^{2}$. Particles at deep potential wells and at interstitial positions are marked blue and green, respectively. The scale bar corresponds to 20$\mu$m.}
  \label{Fig1}
  \end{figure}

Our experiments are performed in a thin sample cell with 200$\mu$m height which is filled with an aqueous suspension of charged polystyrene spheres with diameter $\sigma=1.95 \mu$m. Due to their surface charge they interact via a screened Coulomb potential $\phi(r)\propto\exp(-\kappa r)/r$ where $\kappa^{-1}$ is the Debye screening length which is estimated to about 160 nm. A quasiperiodic light field is created by interfering five laser beams with identical linear polarization and wavelength $\lambda = 532$nm (Fig. \ref{Fig1}(a)). These beams form inside the sample cell a continuous light distribution with decagonal symmetry which imposes optical forces onto the colloidal particles and thus acts as a potential landscape with adjustable potential depth (Fig. \ref{Fig1}(b)). To highlight the quasiperiodic order we have colored potential sites exceeding $50\%$ of the maximum laser intensity in red. Such deep potential wells are aligned along rows which are distributed according to a Fibonacci sequence with separation distances $L=12.1 \mu$m and $S=7.5 \mu$m along the y-direction. The length scales are set by the intersection angle of the laser beams. To avoid buckling effects and vertical particle fluctuations, an additional vertically incident laser beam exerts an optical pressure onto the colloids towards the bottom of the cell. The sample cell is mounted on a xy-piezo stage which allows lateral translations with velocities $\textbf{u}$ in a range of $0.1$ to $10$ $\mu$m/s. Due to viscous Stokes forces  $\textbf{F} = \gamma_{\rm eff}\textbf{u}$ this leads to a lateral driving force onto the colloidal monolayer. The effective drag coefficient $\gamma_{\rm eff}$ is obtained from the short-time particle diffusion coefficient and exploiting the Einstein relation \cite{Bohlein2012}. All experiments were carried out at a particle density slightly below the crystallization threshold where the colloids form a fluid in the absence of the optical landscape and for $\textbf{F} = 0$.
\begin{figure}[htbp]
      \includegraphics[width=0.45\textwidth]{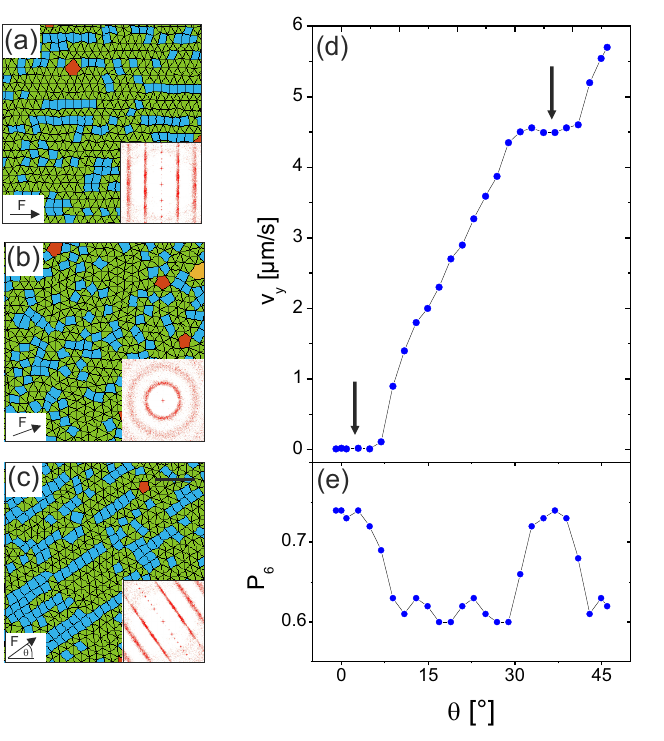}
  \caption{(color online). (a)-(c) Tiling of particle configurations with squares and triangles in blue and green for $\theta = 0^\circ$ (a), $20^\circ$ (b), $36^\circ$ (c). For (a) and (c) tiles arrange in bands which are aligned in the direction of {\bf F} while no such ordering is found in (b). The insets show the structure factor $S(k)$ which displays smectic ordering (a,c) and a moving fluid (b), respectively. The scale bar corresponds to $20 \mu$m. (d) The y-component of the average velocity as a function of $\theta$ exhibits pronounced plateaus at $\theta=0^\circ$ and $36^\circ$ (arrows) demonstrating directional locking. (e) Fraction of sixfold coordinated particles $P_{6}$ vs $\theta$.}
  \label{Fig2}
  \end{figure}

Figure \ref{Fig2} shows how the structure of a colloidal monolayer changes when a driving force ${\bf F}$ is applied at different angles $\theta$ with respect to the symmetry directions of the quasiperiodic substrate (dashed lines in Fig. \ref{Fig1}(b)). We have chosen values of $\theta = 0^\circ$, $20^\circ$ and $36^\circ$, i.e. the first and the last angle corresponding to a symmetry direction and one in between. The particle density was adjusted to $\rho=0.023/\mu m^{2}$ leading to a filling fraction $\eta\approx 2$ being defined by the ratio of deep potential sites (red dots in Fig. \ref{Fig1}b) and the total particle number (Fig. \ref{Fig1}c). For the chosen laser intensity $I = 6.4 \mu W/\mu m^2$, the monolayer forms a quasicrystal for ${\bf F} = 0$ as inferred from the tenfold symmetry in the structure factor (data not shown). For a driving force $F=291$fN, all particles become mobile and travel with rather constant velocity across the potential landscape. In order to visualize the structure of the moving colloidal monolayer, we have applied a tiling algorithm which is based on a Delaunay triangulation of snapshots of the particle positions (for details see \cite{Mikhael2011}). Because these tilings are only subjected to little fluctuations when the monolayer moves across the quasiperiodic substrate, in the following we discuss configurational snapshots rather than time-averaged quantities.

Within the tiling representation, particle positions correspond to the vertices of the tiling pattern shown in Figs. \ref{Fig2}(a)-(c). Obviously, for  $\theta = 0^\circ$ and $36^\circ$ (Figs. \ref{Fig2}(a),(c)) the square and triangular tiles are aligned in bands which are locked to a symmetry direction of the substrate. The corresponding structure factor (insets of Figs. \ref{Fig2}(a),(c)) suggests a
smectic-like phase with high periodic order along the bands and only little periodicity perpendicular to them. In contrast, no such dynamical ordering is found when the direction of ${\bf F}$ deviates from symmetry directions as shown exemplarily for $\theta=20^\circ$ in Fig. \ref{Fig2}(b). Here, the ring structure of $S(k)$ indicates a liquid colloidal state having only short-range order.

\begin{figure}[htbp]
      \includegraphics[width=0.4\textwidth]{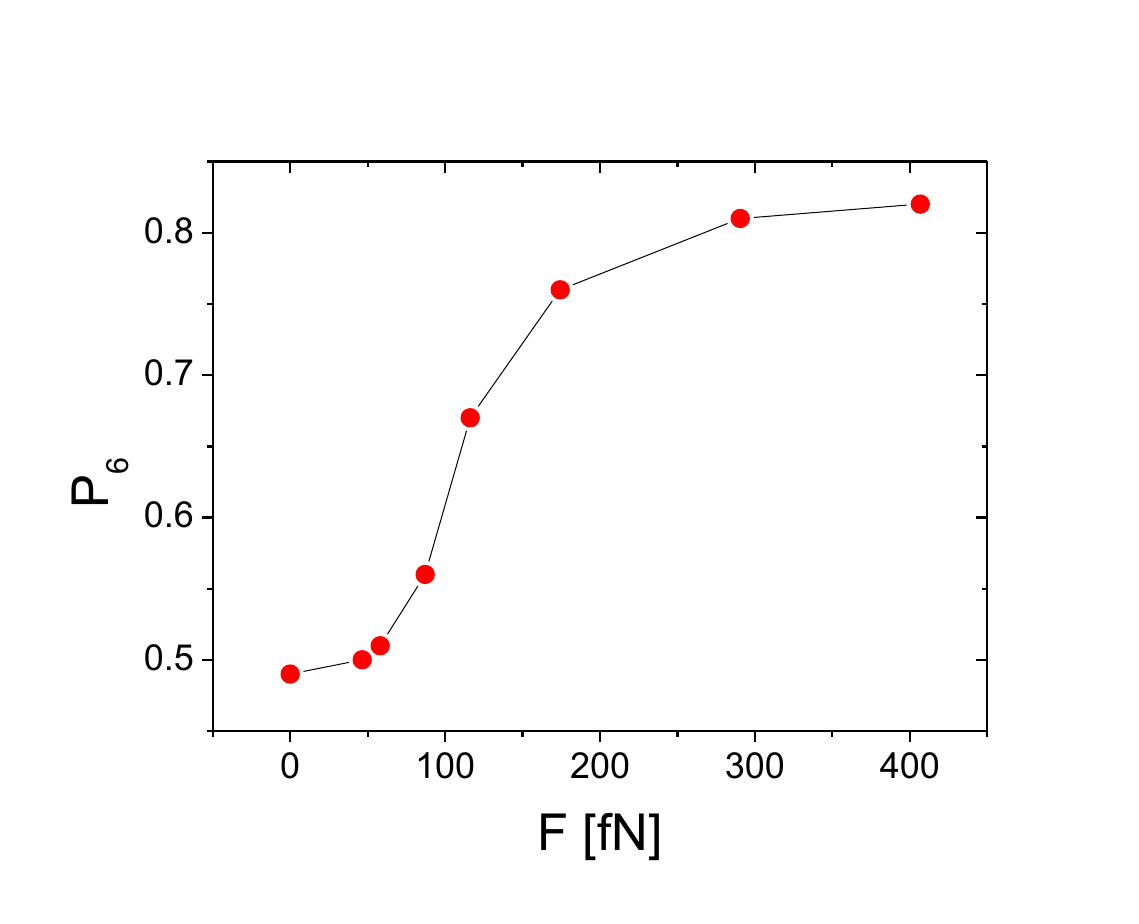}
  \caption{(color online). Fraction of sixfold coordinated particles $P_{6}$ as as function of the driving force for $\theta = 0^\circ$. }
  \label{Fig3}
  \end{figure}

The formation of dynamically ordered phases is associated with discrete locking steps which can be identified by the mean colloidal velocity along the y direction $<v_{y}>=1/N\sum_{1}^{N}\mathbf{v_{i}}\cdot \hat{\mathbf{y}}$ as a function of the driving angle $\theta$ (Fig. \ref{Fig2}(d)).
Instead of a monotonic increase we observe pronounced plateaus (arrows) around  $\theta = 0^\circ$ and $36^\circ$ where the particle's motion becomes locked to the structure of the underlying substrate. The directional locking extends over
an angular range of approximately $10^\circ$ and is centered around multiples of $360^\circ/10$, reflecting the tenfold rotational symmetry of the substrate. To demonstrate the higher degree of order on the locking steps we also determined the fraction of sixfold coordinated particles $P_{6}=\sum_{i}^{N}\delta(z_{i} -6)$ as a function of the direction of the driving force. As seen in Fig. \ref{Fig2}(e), $P_{6}$ peaks at the position of the plateaus which demonstrates a higher amount of periodic order in the locked-in states in correspondence to Figs. \ref{Fig2}(a),(c). In addition to the angle dependence, we also studied how $P_{6}$ depends on the amplitude of the driving force. For $\theta=0^\circ$ we observe a continuous increase and the onset of a saturation value above $F\geq290$fN (Fig. \ref{Fig3}).

Our results are in good agreement with recent numerical studies and support the idea that long-range order rather than translational symmetry is the essential ingredient for locked-in transport \cite{Reichhardt2011}. Contrary to our experiments on a continuous quasiperiodic lattice, the simulations were performed on substrates composed of discrete pinning sites distributed according to a Penrose tiling. This difference may account for the absence of a square phase in our experiments which has been observed in the simulations \cite{Reichhardt2011}.

\begin{figure}[htbp]
      \includegraphics[width=0.4\textwidth]{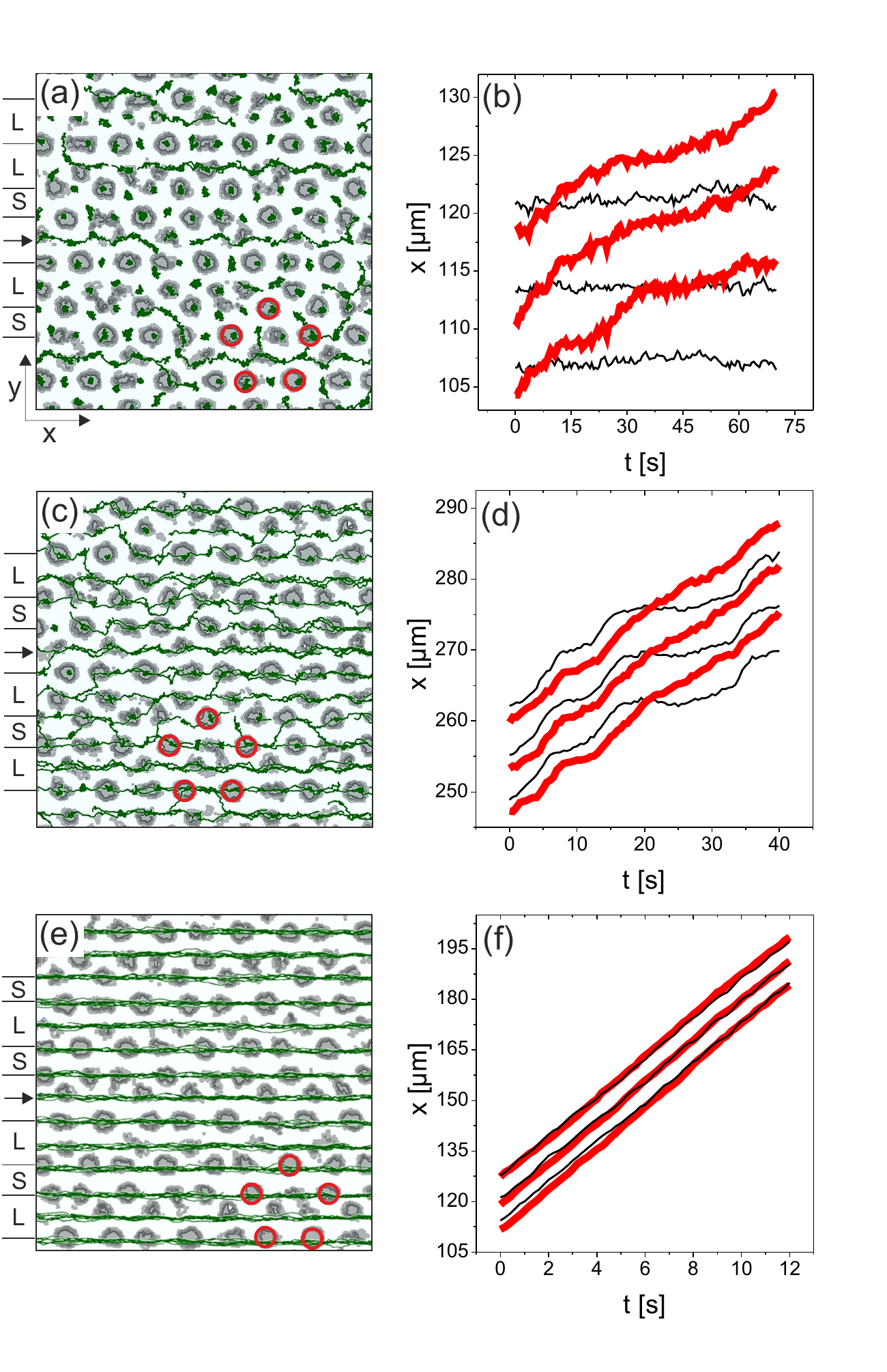}
  \caption{(color online). (a),(c),(e) Particle trajectories superimposed to the quasiperiodic potential for $F=47$fN (a), $F=87$fN (c), $F=291$fN (e) applied along one of the symmetry directions ($\theta=0^\circ$). Deep potential minima are aligned along symmetry axes and form a Fibonacci sequence of lines along the y-direction. Interstitial particles move on lanes along rows of shallow potential minima lying between two L-spaced lines (see arrows). Characteristic motifs are marked with red circles, the laser intensity was set to $I = 6.4 \mu W/\mu m^2$. (b),(d),(f) x-coordinate vs time for three successive trajectories of two adjacent interstitial and non-interstitial lanes. Interstitial and non-interstitial trajectories are marked with thick (red) and thin (black) lines.}
  \label{Fig4}
  \end{figure}
\begin{figure*}[ht]
      \includegraphics[width=0.75\textwidth]{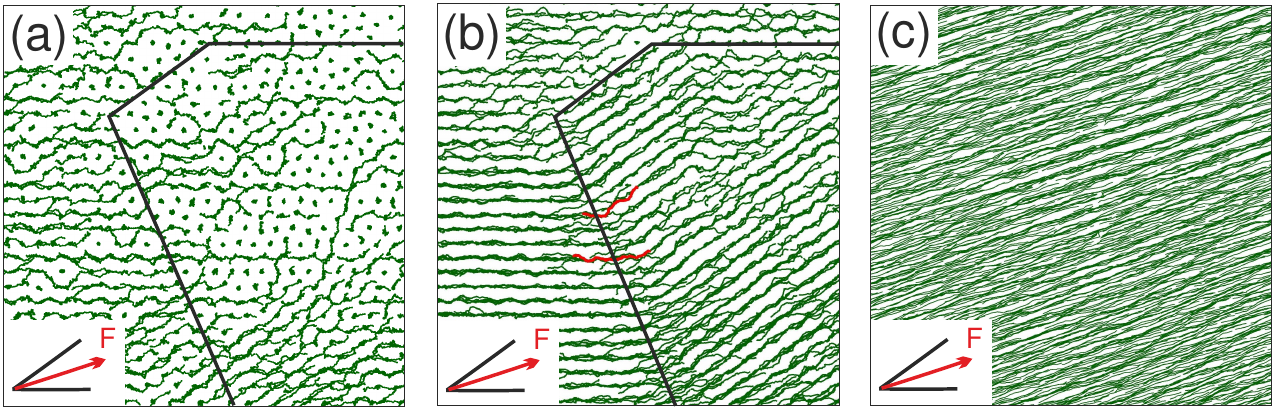}
  \caption{(color online). Particle trajectories for $\theta=18^\circ$ and $F=87$fN (a), $F=115$fN (b), $F=174$fN (c). The light intensity of the interference pattern is $I = 6.4 \mu W/\mu m^{2}$. The insets show the direction of the driving force relative to two symmetry axis.}
  \label{Fig5}
  \end{figure*}
To understand the origin of dynamical ordering on a microscopic level, in  Fig. \ref{Fig4} we show particle trajectories for $F=47, 87, 291$fN and $\theta=0^\circ$. The underlying quasiperiodic substrate potential is plotted as a grey scale background. For $F=47$fN, which is slightly above the depinning threshold, particles at deep substrate pinning sites remain localized while weakly pinned interstitial particles move in the direction of the driving force (Fig. \ref{Fig4}(a)). As a result, trajectories of interstitial particles form lanes in the direction of $\bf F$. These lanes develop predominantly between lines of deep substrate potential wells separated by the distance $L$, because there the substrate is rather shallow (see Fig. \ref{Fig1}(b) and arrows in Fig. \ref{Fig4} (a),(c),(e)). The trajectories x-component of interstitial and non-interstitial particles located in adjacent lanes are plotted in Fig. \ref{Fig4}(b) as thick red and thin black lines, respectively. When the driving force is further increased to $F=87$fN, also previously pinned particles are set into motion and the entire monolayer moves in the direction of $\bf F$ (Fig. \ref{Fig4}(c)). However, interstitial and non-interstitial particles behave differently as seen in Fig. \ref{Fig4}(d). While interstitial particles move with rather constant velocity across the surface, non-interstitial particles show a pronounced stick-slip motion due to their interaction with the underlying deep substrate potential wells. As a result of the additional electrostatic pair potential, the motion of particles becomes correlated along the y-direction which then leads to a coupling between interstitial and non-interstitial lanes.
Further increasing the driving force finally leads to the above described dynamically ordered phase. Fig. \ref{Fig4}(e) shows the trajectories for $F=291$fN. Here, the velocities of non-interstitial and interstitial particles become identical and the entire system moves with rather constant velocity across the substrate which then leads to the smectic-like phase observed above (Fig. \ref{Fig4}(f)).

Dynamical ordering is most pronounced when the particle densities within lanes composed of interstitial and non-interstitial particles are almost identical as this is the case for $\eta \approx 2$ as shown above. Under these conditions neighboring lanes have the same effective interaction which favors a rather equidistant lane spacing. This is achieved by a small shift of particle lanes out of the substrate potential well centers and thus leads to an average lane distance of $d=6.4 \mu m$ (see Fig. \ref{Fig4}(e)).
Because particle exchange between lanes is strongly suppressed, this further supports the formation of dynamically ordered structures.
In general, the formation of a dynamically ordered phase has been predicted to strongly depend on the filling fraction $\eta$ \cite{Reichhardt2011}. This is in good agreement with our
observations that for e.g. $\eta = 1.28$ only disordered phases are observed in our experiments.

Finally, we also investigated the force dependence on the particle trajectories when $\bf F$ deviates from a symmetry direction, i.e. for $\theta=18^\circ$. For $F=87$fN mainly interstitial particles are mobile and form lanes which are aligned along substrate symmetry directions $\theta=0^\circ$, $36^\circ$ and $72^\circ$ Fig. \ref{Fig5}(a)). A further increase to $F=115$fN leads to spatially extended domains (composed of interstitial and non interstitial particles) having different lane alignments $\theta=0^\circ$ and $36^\circ$(Fig. \ref{Fig5}(b)). The orientation of these domains is obviously determined by the direction of the previously, i.e. at lower $\bf F$, formed lanes of interstitial particles which couple to the motion of the non-interstitials (see domain boundaries marked in Figs. \ref{Fig5}(a),(b)). The formation of such dynamic domains requires the sudden change of the direction of particle trajectories as this is seen in our experiments (highlighted trajectories in (Fig. \ref{Fig5}(b)). Such symmetry breaking phenomena in driven monolayers on ordered surfaces have been also predicted on periodic substrate potentials where the entire monolayer will lock to one of the symmetry directions \cite{Reichhardt2004}. This is in contrast to our experiments, where the monolayer partitions into domains which are locked to different angles. We assume that this is due to the higher rotational symmetry of quasiperiodic vs. periodic surfaces which requires smaller kink angles in the particle trajectories.
Finally, for $F=174$fN directional locking is lost and all colloids follow the direction of the applied force (Fig. \ref{Fig5}(c)).

In conclusion we have experimentally investigated the behaviour of a monolayer of interacting particles driven across a substrate with quasiperiodic symmetry. We find that directional locking is not limited to periodic substrates, but also occurs on quasiperiodic potentials which lack translational symmetry. In addition, we demonstrate that the microscopic origin of dynamical ordering transitions is due to the interaction of particle lanes formed by interstitial and non-interstitial particles.

\begin{acknowledgments}
We thank Jules Mikhael and Valentin Blickle for helpful discussions. This work is financially supported by the Deutsche Forschungsgemeinschaft BE 1788/10.
\end{acknowledgments}


\begin{thebibliography}{19}
\expandafter\ifx\csname natexlab\endcsname\relax\def\natexlab#1{#1}\fi
\expandafter\ifx\csname bibnamefont\endcsname\relax
  \def\bibnamefont#1{#1}\fi
\expandafter\ifx\csname bibfnamefont\endcsname\relax
  \def\bibfnamefont#1{#1}\fi
\expandafter\ifx\csname citenamefont\endcsname\relax
  \def\citenamefont#1{#1}\fi
\expandafter\ifx\csname url\endcsname\relax
  \def\url#1{\texttt{#1}}\fi
\expandafter\ifx\csname urlprefix\endcsname\relax\def\urlprefix{URL }\fi
\providecommand{\bibinfo}[2]{#2}
\providecommand{\eprint}[2][]{\url{#2}}

\bibitem[{\citenamefont{Korda et~al.}(2002)\citenamefont{Korda, Taylor, and
  Grier}}]{Korda2002}
\bibinfo{author}{\bibfnamefont{P.~T.} \bibnamefont{Korda}},
  \bibinfo{author}{\bibfnamefont{M.~B.} \bibnamefont{Taylor}},
  \bibnamefont{and} \bibinfo{author}{\bibfnamefont{D.~G.} \bibnamefont{Grier}},
  \bibinfo{journal}{Phys. Rev. Lett.} \textbf{\bibinfo{volume}{89}},
  \bibinfo{pages}{128301} (\bibinfo{year}{2002}).

\bibitem[{\citenamefont{Roichman et~al.}(2007)\citenamefont{Roichman, Wong, and
  Grier}}]{Roichman2007}
\bibinfo{author}{\bibfnamefont{Y.}~\bibnamefont{Roichman}},
  \bibinfo{author}{\bibfnamefont{V.}~\bibnamefont{Wong}}, \bibnamefont{and}
  \bibinfo{author}{\bibfnamefont{D.~G.} \bibnamefont{Grier}},
  \bibinfo{journal}{Phys. Rev. E} \textbf{\bibinfo{volume}{75}},
  \bibinfo{pages}{011407} (\bibinfo{year}{2007}).

\bibitem[{\citenamefont{Pierre-Louis}(2001)}]{PL2001}
\bibinfo{author}{\bibfnamefont{O.}~\bibnamefont{Pierre-Louis}},
  \bibinfo{journal}{Phys. Rev. Lett.} \textbf{\bibinfo{volume}{87}},
  \bibinfo{pages}{106104} (\bibinfo{year}{2001}).

\bibitem[{\citenamefont{Brown et~al.}(1984)\citenamefont{Brown, Mozurkewich,
  and Gruner}}]{Brown1984}
\bibinfo{author}{\bibfnamefont{S.~E.} \bibnamefont{Brown}},
  \bibinfo{author}{\bibfnamefont{G.}~\bibnamefont{Mozurkewich}},
  \bibnamefont{and} \bibinfo{author}{\bibfnamefont{G.}~\bibnamefont{Gruner}},
  \bibinfo{journal}{Phys. Rev. Lett.} \textbf{\bibinfo{volume}{52}},
  \bibinfo{pages}{2277} (\bibinfo{year}{1984}).

\bibitem[{\citenamefont{Reichhardt et~al.}(1997)\citenamefont{Reichhardt,
  Olson, and Nori}}]{Reichhardt1997}
\bibinfo{author}{\bibfnamefont{C.}~\bibnamefont{Reichhardt}},
  \bibinfo{author}{\bibfnamefont{C.~J.} \bibnamefont{Olson}}, \bibnamefont{and}
  \bibinfo{author}{\bibfnamefont{F.}~\bibnamefont{Nori}},
  \bibinfo{journal}{Phys. Rev. Lett.} \textbf{\bibinfo{volume}{78}},
  \bibinfo{pages}{2648} (\bibinfo{year}{1997}).

\bibitem[{\citenamefont{Reichhardt and Nori}(1999)}]{Reichhardt1999}
\bibinfo{author}{\bibfnamefont{C.}~\bibnamefont{Reichhardt}} \bibnamefont{and}
  \bibinfo{author}{\bibfnamefont{F.}~\bibnamefont{Nori}},
  \bibinfo{journal}{Phys. Rev. Lett.} \textbf{\bibinfo{volume}{82}},
  \bibinfo{pages}{414} (\bibinfo{year}{1999}).

\bibitem[{\citenamefont{Morgan and Ketterson}(1998)}]{Morgan1998}
\bibinfo{author}{\bibfnamefont{D.~J.} \bibnamefont{Morgan}} \bibnamefont{and}
  \bibinfo{author}{\bibfnamefont{J.~B.} \bibnamefont{Ketterson}},
  \bibinfo{journal}{Phys. Rev. Lett.} \textbf{\bibinfo{volume}{80}},
  \bibinfo{pages}{3614} (\bibinfo{year}{1998}).

\bibitem[{\citenamefont{MacDonald et~al.}(2003)\citenamefont{MacDonald,
  Spalding, and Dholakia}}]{MacDonald2003}
\bibinfo{author}{\bibfnamefont{M.~P.} \bibnamefont{MacDonald}},
  \bibinfo{author}{\bibfnamefont{G.~C.} \bibnamefont{Spalding}},
  \bibnamefont{and} \bibinfo{author}{\bibfnamefont{K.}~\bibnamefont{Dholakia}},
  \bibinfo{journal}{Nature} \textbf{\bibinfo{volume}{426}},
  \bibinfo{pages}{421} (\bibinfo{year}{2003}).

\bibitem[{\citenamefont{Speer et~al.}(2009)\citenamefont{Speer, Eichhorn, and
  Reimann}}]{Speer2009}
\bibinfo{author}{\bibfnamefont{D.}~\bibnamefont{Speer}},
  \bibinfo{author}{\bibfnamefont{R.}~\bibnamefont{Eichhorn}}, \bibnamefont{and}
  \bibinfo{author}{\bibfnamefont{P.}~\bibnamefont{Reimann}},
  \bibinfo{journal}{Phys. Rev. Lett.} \textbf{\bibinfo{volume}{102}},
  \bibinfo{pages}{124101} (\bibinfo{year}{2009}).

\bibitem[{\citenamefont{Reichhardt and Reichhardt}(to be
  published)}]{Reichhartunpublished}
\bibinfo{author}{\bibfnamefont{C.}~\bibnamefont{Reichhardt}} \bibnamefont{and}
  \bibinfo{author}{\bibfnamefont{C.~J.~O.} \bibnamefont{Reichhardt}},
  \bibinfo{journal}{Journal of Physics: Cond. Mat.}  (\bibinfo{year}{to be
  published}).

\bibitem[{\citenamefont{Kemmler et~al.}(2006)\citenamefont{Kemmler, Gurlich,
  Sterck, Pohler, Neuhaus, Siegel, Kleiner, and Koelle}}]{Kemmler2006}
\bibinfo{author}{\bibfnamefont{M.}~\bibnamefont{Kemmler}},
  \bibinfo{author}{\bibfnamefont{C.}~\bibnamefont{Gurlich}},
  \bibinfo{author}{\bibfnamefont{A.}~\bibnamefont{Sterck}},
  \bibinfo{author}{\bibfnamefont{H.}~\bibnamefont{Pohler}},
  \bibinfo{author}{\bibfnamefont{M.}~\bibnamefont{Neuhaus}},
  \bibinfo{author}{\bibfnamefont{M.}~\bibnamefont{Siegel}},
  \bibinfo{author}{\bibfnamefont{R.}~\bibnamefont{Kleiner}}, \bibnamefont{and}
  \bibinfo{author}{\bibfnamefont{D.}~\bibnamefont{Koelle}},
  \bibinfo{journal}{Phys. Rev. Lett.} \textbf{\bibinfo{volume}{97}},
  \bibinfo{pages}{147003} (\bibinfo{year}{2006}).

\bibitem[{\citenamefont{Silhanek et~al.}(2006)\citenamefont{Silhanek, Gillijns,
  Moshchalkov, Zhu, Moonens, and Leunissen}}]{Silhanek2006}
\bibinfo{author}{\bibfnamefont{A.~V.} \bibnamefont{Silhanek}},
  \bibinfo{author}{\bibfnamefont{W.}~\bibnamefont{Gillijns}},
  \bibinfo{author}{\bibfnamefont{V.~V.} \bibnamefont{Moshchalkov}},
  \bibinfo{author}{\bibfnamefont{B.~Y.} \bibnamefont{Zhu}},
  \bibinfo{author}{\bibfnamefont{J.}~\bibnamefont{Moonens}}, \bibnamefont{and}
  \bibinfo{author}{\bibfnamefont{L.~H.~A.} \bibnamefont{Leunissen}},
  \bibinfo{journal}{Appl. Phys. Lett.} \textbf{\bibinfo{volume}{89}},
  \bibinfo{pages}{152507} (\bibinfo{year}{2006}).

\bibitem[{\citenamefont{Koshelev and Vinokur}(1994)}]{Koshelev1994}
\bibinfo{author}{\bibfnamefont{A.~E.} \bibnamefont{Koshelev}} \bibnamefont{and}
  \bibinfo{author}{\bibfnamefont{V.~M.} \bibnamefont{Vinokur}},
  \bibinfo{journal}{Physical Review Letters} \textbf{\bibinfo{volume}{73}},
  \bibinfo{pages}{3580} (\bibinfo{year}{1994}).

\bibitem[{\citenamefont{Pardo et~al.}(1998)\citenamefont{Pardo, de~la Cruz,
  Gammel, Bucher, and Bishop}}]{Pardo1998}
\bibinfo{author}{\bibfnamefont{F.}~\bibnamefont{Pardo}},
  \bibinfo{author}{\bibfnamefont{F.}~\bibnamefont{de~la Cruz}},
  \bibinfo{author}{\bibfnamefont{P.~L.} \bibnamefont{Gammel}},
  \bibinfo{author}{\bibfnamefont{E.}~\bibnamefont{Bucher}}, \bibnamefont{and}
  \bibinfo{author}{\bibfnamefont{D.~J.} \bibnamefont{Bishop}},
  \bibinfo{journal}{Nature} \textbf{\bibinfo{volume}{396}},
  \bibinfo{pages}{348} (\bibinfo{year}{1998}).

\bibitem[{\citenamefont{Reichhardt and Reichhardt}(2011)}]{Reichhardt2011}
\bibinfo{author}{\bibfnamefont{C.}~\bibnamefont{Reichhardt}} \bibnamefont{and}
  \bibinfo{author}{\bibfnamefont{C.~J.~O.} \bibnamefont{Reichhardt}},
  \bibinfo{journal}{Phys. Rev. Lett.} \textbf{\bibinfo{volume}{106}},
  \bibinfo{pages}{060603} (\bibinfo{year}{2011}).

\bibitem[{\citenamefont{Mikhael et~al.}(2008)\citenamefont{Mikhael, Roth,
  Helden, and Bechinger}}]{Mikhael2008}
\bibinfo{author}{\bibfnamefont{J.}~\bibnamefont{Mikhael}},
  \bibinfo{author}{\bibfnamefont{J.}~\bibnamefont{Roth}},
  \bibinfo{author}{\bibfnamefont{L.}~\bibnamefont{Helden}}, \bibnamefont{and}
  \bibinfo{author}{\bibfnamefont{C.}~\bibnamefont{Bechinger}},
  \bibinfo{journal}{Nature} \textbf{\bibinfo{volume}{454}},
  \bibinfo{pages}{501} (\bibinfo{year}{2008}).

\bibitem[{\citenamefont{Bohlein et~al.}(2012)\citenamefont{Bohlein, J., and
  Bechinger}}]{Bohlein2012}
\bibinfo{author}{\bibfnamefont{T.}~\bibnamefont{Bohlein}},
  \bibinfo{author}{\bibfnamefont{M.}~\bibnamefont{J.}}, \bibnamefont{and}
  \bibinfo{author}{\bibfnamefont{C.}~\bibnamefont{Bechinger}},
  \bibinfo{journal}{Nature Materials} \textbf{\bibinfo{volume}{11}},
  \bibinfo{pages}{126} (\bibinfo{year}{2012}).

\bibitem[{\citenamefont{Mikhael et~al.}(2011)\citenamefont{Mikhael, Gera,
  Bohlein, and Bechinger}}]{Mikhael2011}
\bibinfo{author}{\bibfnamefont{J.}~\bibnamefont{Mikhael}},
  \bibinfo{author}{\bibfnamefont{G.}~\bibnamefont{Gera}},
  \bibinfo{author}{\bibfnamefont{T.}~\bibnamefont{Bohlein}}, \bibnamefont{and}
  \bibinfo{author}{\bibfnamefont{C.}~\bibnamefont{Bechinger}},
  \bibinfo{journal}{Soft Matter} \textbf{\bibinfo{volume}{7}},
  \bibinfo{pages}{1352} (\bibinfo{year}{2011}).

\bibitem[{\citenamefont{Reichhardt and Reichhardt}(2004)}]{Reichhardt2004}
\bibinfo{author}{\bibfnamefont{C.}~\bibnamefont{Reichhardt}} \bibnamefont{and}
  \bibinfo{author}{\bibfnamefont{C.~J.~O.} \bibnamefont{Reichhardt}},
  \bibinfo{journal}{EPL} \textbf{\bibinfo{volume}{68}},
  \bibinfo{pages}{303} (\bibinfo{year}{2004}).

\end{thebibliography}
\end{document}